# Searching for Scientific Evidence in a Pandemic: An Overview of TREC-COVID

Kirk Roberts, Tasmeer Alam, Steven Bedrick, Dina Demner-Fushman, Kyle Lo, Ian Soboroff, Ellen Voorhees, Lucy Lu Wang, William R Hersh

Corresponding Author
Kirk Roberts, kirk.roberts@uth.tmc.edu

Affiliations
Kirk Roberts
University of Texas Health Science Center at Houston, Houston, Texas, USA

Tasmeer Alam
Morgan State University, Baltimore, Maryland, USA

Steven Bedrick
Oregon Health & Science University, Portland, Oregon, USA

Dina Demner-Fushman
US National Library of Medicine, Bethesda, Maryland, USA

Kyle Lo
Allen Institute for AI, Seattle, Washington, USA

Ian Soboroff
National Institute of Standards and Technology, Gaithersburg, Maryland, USA

Ellen Voorhees
National Institute of Standards and Technology, Gaithersburg, Maryland, USA

Lucy Lu Wang
Allen Institute for AI, Seattle, Washington, USA

William R Hersh
Oregon Health & Science University, Portland, Oregon, USA

**ABSTRACT**

We present an overview of the TREC-COVID Challenge, an information retrieval (IR) shared task to evaluate search on scientific literature related to COVID-19. The goals of TREC-COVID include the construction of a pandemic search test collection and the evaluation of IR methods for COVID-19. The challenge was conducted over five rounds from April to July, 2020, with participation from 92 unique teams and 556 individual submissions. A total of 50 topics (sets of related queries) were used in the evaluation, starting at 30 topics for Round 1 and adding 5 new topics per round to target emerging topics at that state of the still-emerging pandemic. This paper provides a comprehensive overview of the structure and results of TREC-COVID. Specifically, the paper provides details on the background, task structure, topic structure, corpus, participation, pooling, assessment, judgments, results, top-performing systems, lessons learned, and benchmark datasets.

# 1. INTRODUCTION

The Coronavirus Disease 2019 (COVID-19) pandemic has resulted in an enormous demand for and supply of evidence-based information. On the demand side, there are numerous information needs regarding the basic biology, clinical treatment, and public health response to COVID-19. On the supply side, there have been a vast number of scientific publications, including preprints. Despite the large supply of available scientific evidence, beyond the medical aspects of the pandemic, COVID-19 has resulted in an "infodemic" as well [1-3] with large amounts of confusion, disagreement, and distrust about available information.

A key component in identifying available evidence is by accessing the scientific literature using the best possible information retrieval (IR, or search) systems. As such, there was a need for rapid implementation of IR systems tuned for such an environment and a comparison of the efficacy of those systems. A common approach for large-scale comparative evaluation of IR systems is the challenge evaluation, with the largest and best-known approach coming from the Text Retrieval Conference (TREC) organized by the US National Institute of Standards and Technology (NIST) [4]. The TREC framework was applied to the COVID-19 Open Research Dataset (CORD-19), a dynamic resource of scientific papers on COVID-19 and related historical coronavirus research [5].

The primary goal of the TREC-COVID Challenge was to build a test collection for evaluating search engines dealing with the complex information landscape in events such as a pandemic. Since IR focuses on large document collections and it is infeasible to manually judge every document for every topic, IR test collections are generally built via manual judgment using participants' retrieval results to guide the selection of which documents to judge. This allows for a wide variety of search techniques to identify potentially relevant documents, and focuses the manual effort on just those documents most likely to be relevant. Thus, to build an excellent test collection for pandemics, it is necessary to conduct a shared task such as TREC-COVID with a large, diverse set of participants.

A critical aspect of a pandemic is the temporal nature of the event: as new information arises a search engine must adapt to these changes, including the rapid pace with which new discoveries are added to the growing corpus of scientific knowledge on the pandemic. The three distinct aspects of temporality in the context of the pandemic are (1) rapidly changing information needs: as knowledge about the pandemic grows, the information needs evolve to include both the new aspects of the existing topics and new topics; (2) rapidly changing state of knowledge reflected both in the high rate at which the new work is published and the initial publications are edited; and (3) heterogeneity of the relevant work: whereas in traditional biomedical collections the documents and journals are peer-reviewed, in a pandemic scenario any publications, e.g., preprints, containing new information may be relevant and may actually contain the most up-to-date information. The result of all these factors is that the best search strategy at the beginning of a pandemic (with small amounts of scattered information, many unknowns) may be different than the best strategy mid-pandemic (rapidly-growing burst of information with some emerging answers, unknowns still exist but are better defined) or after the pandemic (many more answers but with a corpus that contains a significant evolution through time, may require filtering out many of the early pandemic information that has become outdated). TREC-COVID models the pandemic stage using a multi-round structure, where more documents are available and additional topics are added as new questions emerge.

The other critical aspect of a pandemic from an IR perspective is the ability to gather feedback on search performance as a pandemic proceeds. As new topics emerge, judgments on these topics can be collected (manually or automatically, e.g. click data) that can be used to improve search performance (both on that topic of interest and other topics). This is subject to similar temporality constraints as above: feedback is only available on documents that previously exist, while the amount of feedback data available steadily grows over the course of the pandemic.

These two aspects—temporality of data and the availability of relevance feedback for model development—are the two core contributions of TREC-COVID from an IR perspective. From a biomedical perspective, TREC-COVID's contributions include its unique focus on an emerging infectious disease, the inclusion of both peer-reviewed and preprint articles, and its substantial size in terms of the number of judgments and proportion of the collection that was judged. Finally, a practical contribution of TREC-COVID was the rapid availability of its manual judgments so that public-facing COVID-19-focused search engines could tune their approach to best help researchers and consumers find evidence in the midst of the pandemic.

TREC-COVID was structured as a series of rounds to capture these changes. Over five rounds of evaluation, TREC-COVID received 556 submissions with 92 participating teams. The final test collection contains 69,318 manual judgments on 50 topics important to COVID-19. Each round included an increasing number of topics pertinent to the pandemic, where each topic is a set of queries around a common theme (e.g., dexamethasone) provided at three levels of granularity (described in Section 4). Capturing the evolving corpus proved to be quite challenging as preprints were released, updated, and published, sometimes with substantial changes in content. An additional benefit of the multi-round structure of the collection was support of research on *relevance feedback*, supervised machine learning techniques that find additional relevant documents for a topic by exploiting existing relevance judgments.

This paper provides a complete overview of the entire TREC-COVID Challenge. In prior publications, we provided our initial rationale for TREC-COVID and its structure [6] as well as a snapshot of the task after the first round [7]. Additional post-hoc evaluations have also been conducted comparing system qualities [8] and assessing the quality of the final collection [9]. This paper presents a description of the overall challenge now that it has formally concluded. Section 2 places TREC-COVID within the scientific context of IR shared tasks. Section 3 provides an overview of the overall task structure. Section 4 explains the topic structure, how the topics were created, and what types of topics were used. Section 5 details the corpus that systems searched over. Section 6 provides the participation statistics and list of submission information. Section 7 describes how those runs were pooled to select documents for evaluation. Section 8 details the assessment process: who performed the judging, how it was done, and what types of judgments were made. Section 9 describes the resulting judgment sets. Section 10 provides the overall results of the participant systems across the different metrics used in the task. Section 11 contains short descriptions of the systems with published descriptions. Section 12 discusses some of the lessons learned by the TREC-COVID organizers, including lessons for IR research in general, COVID-19 search in particular, and the construction of pandemic test collections should the unfortunate opportunity arise to create another such test collection amidst a new pandemic. Finally, Section 13 describes the different benchmark test collections resulting from TREC-COVID. All data produced during TREC-COVID has been archived on the TREC-COVID web site at http://ir.nist.gov/trec-covid/.

## 2. RELATED WORK

While there has never been an IR challenge evaluation specifically for pandemics, there is a rich history of biomedical IR evaluations, especially within TREC. Similar to TREC-COVID, most of these evaluations have focused on retrieving biomedical literature. The TREC Genomics track (2003-2007) [10-15] targeted biomedical researchers interested in the genomics literature. The TREC Clinical Decision Support track (2014-2016) [16-18] targeted clinicians interested in finding evidence for diagnosis, testing, and treatment of patients. The TREC Precision Medicine track (2017-2020) [19-22] refined that focus to oncologists interested in treating cancer patients with actionable gene mutations. Beyond these, the TREC Medical Records track [23,24] focused on retrieving patient records for building research cohorts (e.g., for clinical trial recruitment). The Medical ImageCLEF tasks [25-28] focused on the multi-modal (text and image) retrieval of medical images (e.g., chest x-rays). Finally, the CLEF eHealth tasks [29,30] focused largely on retrieval for health consumers (patients, caregivers, and other non-medical professionals). TREC-COVID differs from these in terms of medical content, as no prior evaluation had focused on infectious diseases, much less pandemics. However, TREC-COVID also differs from these tasks in terms of its temporal structure, which enables evaluating how search engines adapt to a changing information landscape.

As mentioned earlier, TREC-COVID provided infrastructural support for research on relevance feedback.  Broadly speaking, a relevance feedback technique is any search method that uses known relevant documents to retrieve additional relevant documents for the same topic.  The now-classic "more like this" query is a prototypical relevance feedback search. Information filtering, in which a user's standing information need is used to select the documents in a stream that should be returned to the user, can be cast as a feedback problem in that feedback from the user on documents retrieved earlier in the stream informs the selection of documents later in the stream.  TREC focused research on the filtering task with the Filtering track, and in TREC 2002 track organizers used relevance feedback algorithms to select documents for assessors to judge to create the ground truth data for the track [31]. But the filtering task is a special case of feedback where the emphasis is on the on-line learning of the information need.  Other TREC tracks including the Robust track, Common Core track, and the current Deep Learning track re-used topics from one test collection to target a separate document set. In these tracks the focus has been on the viability of the transfer learning. TREC also included a Relevance Feedback track in TRECs 2008 and 2009 [32] with the explicit goal of creating an evaluation framework for direct comparison of feedback reformulation algorithms. The track created the framework, but it was based on an existing test collection with randomly selected, very small numbers of relevant documents as the test conditions. TREC-COVID also enabled participants to compare feedback techniques using identical relevance sets, but in contrast to the other tracks, these sets were naturally occurring and relatively large, were targeted at the same document set, and contain multiple iterations of feedback.

## 3. TASK STRUCTURE

The standard TREC evaluation involves providing participants with a fixed corpus and a fixed set of topics, as well as having a timeline that lasts several months (2-6 months to submit results, 1-3 months to conduct assessment). As previously described, these constraints are not compatible with pandemic search, since the corpus is constantly growing, topics of interest are constantly emerging, systems need to be built quickly, and assessment needs to occur rapidly. Hence, the structure of a pandemic IR shared task must diverge from the standard TREC model in several important and novel ways.

TREC-COVID was conceived as a multi-round evaluation, where in each round an updated corpus would be used, the number of topics would increase, and participants would submit new results. An initial, somewhat arbitrary, choice of five rounds was proposed to ensure enough iterations to evaluate the temporal aspects of the task while keeping manual assessment feasible. The time between rounds was proposed to be limited to just 2-3 weeks in order to capture rapid snapshots of the state of the pandemic. Ultimately, the task did indeed last five rounds and the iteration format was largely adopted.

[INSERT FIGURE 1 HERE]
Figure 1. High-level structure of TREC-COVID.

A high-level overview of the structure of TREC-COVID is shown in Figure 1. This highlights the interactions between rounds, assessment, and corpus. Table 1 provides the timeline of the task, including the round, start/end dates, release and size of the corpus, number of topics, participation, and cumulative judgments available after the completion of that round. The start date of a round is when the topics were made available as well as the manual judgments from the prior round. The end date is when submissions were due for that round. In between rounds, manual judging occurred for the prior round (referred to below as the *X*.0 judging for Round *X*). During the next round, while participants were developing systems using the Round *X*.0 (and all prior) data, additional manual judging occurred for the prior round (referred to as the *X*.5 judging). However, these would not be available until the conclusion of the next round (Round *X*+1). This enabled a near-constant judging process to maximize the number of manual judgements while still keeping to a rapid iteration schedule.

Table 1. Overview of the TREC-COVID timeline over the five rounds.

|  | **Round 1** | **Round 2** | **Round 3** | **Round 4** | **Round 5** |
|---|---|---|---|---|---|
| **CORD-19 Release** | Apr 10 | May 1 | May 19 | Jun 19 | Jul 16 |
| **Topic Release Date** | Apr 15 | May 4 | May 26 | Jun 26 | Jul 22 |
| **Submission Date** | Apr 23 | May 13 | Jun 3 | Jul 6 | Aug 3 |
| **Corpus Size** (articles) | 51,103 | 59,851 | 128,492 | 157,817 | 191,175 |
| **Topics** | 30 | 35 | 40 | 45 | 50 |
| **Participation** (teams) | 56 | 51 | 31 | 27 | 28 |
| **Participation** (submissions) | 143 | 136 | 79 | 72 | 126 |
| **Manual Judgments** (cumulative) | 8,691 | 20,728 | 33,068 | 46,203 | 69,318 |

As can be seen in Table 1, Round 1 started with 30 topics and 5 new topics were added every round. This allowed for emerging "hot" topics to respond to the evolving nature of the pandemic.

The participation numbers in Table 1 reflect the number of unique teams for each round and the total number of submissions for that round. Teams were restricted to a maximum of three submissions per round except for Round 5 when the limit was eight submissions. The participation numbers include a baseline "team" and several baseline submissions starting in Round 2. The baselines were provided by the University of Waterloo based on the Anserini toolkit [33,34] for the purpose of providing a common yardstick between rounds and to encourage teams to use all three of their submissions for non-baseline methods.

The manual judgment numbers in Table 1 reflect the TREC-COVID test collection grew from quite a small IR test collection in terms of manual judgments to a large collection (smaller than many of the ad hoc TREC tracks in the 1990s, but larger than almost any TREC track since). Critically, the size of an IR test collection can also be measured relative to the corpus size (i.e., what percentage of documents are judged for a given topic), and from this perspective the TREC-COVID test collection is enormous with some topics having 1% of CORD-19 judged (more details on the topics are provided in the next section, while details on the CORD-19 corpus are provided in Section 5). The cumulative numbers include the *X*.0 judgments for that round as well as the *X*.0 and *X*.5 judgments for prior rounds, with the exceptions that articles removed in that version of CORD-19 were removed from the judgments and articles that needed to be re-judged due to updates in CORD-19 are not double-counted. Note that there was an initial Round 0.5 judgment set (based on Anserini runs) but no Round 5.5 judgments. The number of judgments was not strictly based on the number of submissions, as the pooling described in Section 7 allowed for a flexible number of top-ranked articles to be selected for judging. Instead, factors such as timing, funding, and the availability of assessors largely dictated the number of judgments performed for each Round.

## 4. TOPICS

The search topics have a three-part structure, with increasing levels of granularity. The *query* is a few keywords, analogous to most queries submitted to search engines. The *question* is a natural language question that more clearly expresses the information need, and is a more complete alternative to the query. Finally, the *narrative* is a longer exposition of the topic, which provides more details and possible clarifications, but does not necessarily contain all the information provided in the question. Table 2 lists three example topics from different rounds. All topics referred directly to COVID-19 or the SARS-CoV-2 virus, but in some cases the broader term "coronavirus" was used in either the query or question. For some of these topics, background information on other coronaviruses could be partially relevant, but was left to the discretion of the manual assessors. See Voorhees et al. [7] for a more thorough discussion on this terminology issue.

Table 2. Three example TREC-COVID topics.

| **Topic 12** (introduced Round 1) |
| --- |
| Query: coronavirus quarantine |
| Question: what are best practices in hospitals and at home in maintaining quarantine? |
| Narrative: Seeking information on best practices for activities and duration of quarantine for those exposed and/ infected to COVID-19 virus. |
| **Topic 36** (introduced Round 3)<br>Query: SARS-CoV-2 spike structure<br>Question: What is the protein structure of the SARS-CoV-2 spike?<br>Narrative: Looking for studies of the structure of the spike protein on the virus using any methods, such as cryo-EM or crystallography |
| **Topic 46** (introduced Round 5)<br>Query: dexamethasone coronavirus<br>Question: what evidence is there for dexamethasone as a treatment for COVID-19?<br>Narrative: Looking for studies on the impact of dexamethasone treatment in COVID-19 patients, including health benefits as well as adverse effects. This also includes specific populations that are benefitted/harmed by dexamethasone. |

The topics were designed to be responsive to many of the scientific needs of the major stakeholders of the biomedical research community. The topics were intentionally balanced between bench science (e.g., microbiology, proteomics, drug modeling), clinical science (e.g., drug effectiveness in human trials, clinical safety), and public health (e.g., prevention measures, population-level impact of the disease). Soni and Roberts [35] conducted a post-hoc categorization of the first 30 topics along these lines, as well as a separate categorization based on function: whether the topic focused on the transmission of the virus, actions to aid prevention of contracting the disease, the effect of COVID-19 on the body or populations, and treatment efforts.

Several efforts were made to ensure the topics were broadly representative of the needs of the pandemic. Calls were put out via social media asking for community input for topic ideas. Queries submitted to the National Library of Medicine were examined to gauge concerns of the wider public. Additionally, the streams of prominent Twitter medical influencers were examined to identify hot topics in the news. The iterative nature of the task also enabled the topics to adapt to the evolving needs of the pandemic. For every round, five new topics were created in an effort to both address any deficiencies in the existing topics as well as to include recently high-profile topics that received little scientific attention at the time of the prior rounds (e.g., the major dexamethasone trial [36] was not published until July, just in time for Round 5).

Table 3 lists the query for all the topics used in the task, as well as an extension of the Soni & Roberts [35] categories to all 50 topics. Again, these categories were not intended to be authoritative, merely to help balance the types of topics used in the challenge and aid in post-hoc analysis. Many—or even most—topics could feasibly fit into multiple categories. We provide this here for the purpose of providing insights into the types of topics used in the challenge.

Table 3: All 50 topics (only the Query field) along with the research field and function categories assigned to each topic.

| | Topic | Assigned Category | |
|---|---|---|---|
| Number | Query | Research Field | Function |
| 1 | coronavirus origin | Biological | Transmission |
| 2 | coronavirus response to weather changes | Public Health | Transmission |
| 3 | coronavirus immunity | Clinical | Prevention |
| 4 | how do people die from the coronavirus | Clinical | Effect |
| 5 | animal models of COVID-19 | Biological | Treatment |
| 6 | coronavirus test rapid testing | Public Health | Prevention |
| 7 | serological tests for coronavirus | Public Health | Prevention |
| 8 | coronavirus under reporting | Public Health | Prevention |
| 9 | coronavirus in Canada | Public Health | Transmission |
| 10 | coronavirus social distancing impact | Public Health | Prevention |
| 11 | coronavirus hospital rationing | Clinical | Treatment |
| 12 | coronavirus quarantine | Public Health | Prevention |
| 13 | how does coronavirus spread | Biological | Transmission |
| 14 | coronavirus super spreaders | Public Health | Transmission |
| 15 | coronavirus outside body | Biological | Transmission |
| 16 | how long does coronavirus survive on surfaces | Biological | Transmission |
| 17 | coronavirus clinical trials | Clinical | Prevention |
| 18 | masks prevent coronavirus | Public Health | Prevention |
| 19 | what alcohol sanitizer kills coronavirus | Biological | Prevention |
| 20 | coronavirus and ACE inhibitors | Biological | Effect |
| 21 | coronavirus mortality | Public Health | Effect |
| 22 | coronavirus heart impacts | Clinical | Effect |
| 23 | coronavirus hypertension | Clinical | Effect |
| 24 | coronavirus diabetes | Clinical | Effect |
| 25 | coronavirus biomarkers | Biological | Effect |
| 26 | coronavirus early symptoms | Clinical | Effect |
| 27 | coronavirus asymptomatic | Clinical | Transmission |
| 28 | coronavirus hydroxychloroquine | Clinical | Treatment |
| 29 | coronavirus drug repurposing | Biological | Treatment |
| 30 | coronavirus remdesivir | Clinical | Treatment |
| 31 | difference between coronavirus and flu | Biological | N/A |
| 32 | coronavirus subtypes | Biological | N/A |
| 33 | coronavirus vaccine candidates | Clinical | Treatment |
| 34 | coronavirus recovery | Clinical | Effect |

| 35 | coronavirus public datasets | Biological | Transmission |
|---|---|---|---|
| 36 | SARS-CoV-2 spike structure | Biological | Transmission |
| 37 | SARS-CoV-2 phylogenetic analysis | Biological | N/A |
| 38 | COVID inflammatory response | Clinical | Effect |
| 39 | COVID-19 cytokine storm | Biological | Effect |
| 40 | coronavirus mutations | Biological | Transmission |
| 41 | COVID-19 in African-Americans | Public Health | Effect |
| 42 | Vitamin D and COVID-19 | Clinical | Treatment |
| 43 | violence during pandemic | Public Health | Effect |
| 44 | impact of masks on coronavirus transmission | Public Health | Prevention |
| 45 | coronavirus mental health impact | Public Health | Effect |
| 46 | dexamethasone coronavirus | Clinical | Treatment |
| 47 | COVID-19 outcomes in children | Clinical | Effect |
| 48 | school reopening coronavirus | Public Health | Prevention |
| 49 | post-infection COVID-19 immunity | Public Health | Effect |
| 50 | mRNA vaccine coronavirus | Biological | Treatment |

## 5. CORPUS

TREC-COVID uses documents from the COVID-19 Open Research Dataset (CORD-19) [5], a corpus created to support text mining, information retrieval, and natural language processing over the COVID-19 literature. The corpus is released daily by the Allen Institute for AI and partner institutions Chan Zuckerberg Initiative, Georgetown Center for Security and Emerging Technology, IBM Research, Kaggle, Microsoft Research, the National Library of Medicine at NIH, and The White House Office of Science and Technology Policy. CORD-19 was first published on March 16, 2020 with 28K documents, and has grown to include more than 280K entries. In recent months, COVID-19 literature has been published at an unprecedented rate, with several hundred new papers being released each day, challenging the ability of clinicians, researchers, and policymakers to keep up with the latest research.

The CORD-19 corpus aims to support automated systems for literature search, discovery, exploration, and summarization that help to address issues of information overload. The corpus includes papers and preprints on COVID-19 and historical coronaviruses, sourced from PubMed Central, PubMed, bioRxiv, medRxiv, arXiv, the WHO's COVID-19 database, Semantic Scholar, and more. Documents are selected based on the presence of a set of keywords associated with the coronavirus family—including *COVID*, *COVID-19*, *Coronavirus*, *Corona virus*, *2019-nCoV*, *SARS-CoV*, *MERS-CoV*, *Severe Acute Respiratory Syndrome*, and *Middle East Respiratory*

*Syndrome*—in the title, abstract, or body text of the document. Each document in the corpus is associated with normalized document metadata; for open access documents, structured full text is extracted using the S2ORC pipeline [37] and made available as part of the corpus. CORD-19 performs simple deduplication over source documents. The dataset takes a conservative deduplication approach; documents are merged into a single entry if and only if they share at least one of the following identifiers in common: DOI, PubMed ID, PMC ID, and arXiv ID, or have the same title, while having no conflicts between identifiers. Though this method is able to identify obvious duplicates, it does not address the merging of similar but non-identical documents, e.g., a preprint and its ultimate publication. In these cases, we choose not to merge the preprint and publication. Since preprints can undergo significant changes prior to publication, we believe this choice is justified. However, for retrieval, additional deduplication may be necessary.

Another unique feature of CORD-19 is that it is updated daily, an attempt to keep up with the hundreds of new papers released everyday. Each round in TREC-COVID is anchored to a specific release of CORD-19 (as shown in Table 1), with the corpus growing from 47K documents in April for Round 1 to 191K documents in July for Round 5. CORD-19 attempts to provide stable identifiers (CORD UID) across different versions of the dataset. This is accomplished by aligning each document in a particular release with identical documents in the prior release based on shared document identifiers. In general, this method performs well. However, issues in automated CORD-19 corpus generation have caused partial loss of persistence between neighboring versions. To offset this issue, TREC-COVID provides identifier mappings between versions of CORD-19 used in the shared task. These files identify documents which differ on CORD UID between TREC dataset versions but are nonetheless the same document based on other unique identifiers and/or titles.

The majority of the documents in CORD-19 were published in 2020 and are on the subject of COVID-19. Around a quarter of the articles are in the field of virology, followed by articles on the medical specialties of immunology, surgery, internal medicine, and intensive care medicine, as classified by Microsoft Academic fields of study [38]. The corpus has been used by clinical researchers as a source of documents for systematic literature reviews on COVID-19, and has been the foundation of many dozens of search and exploration interfaces for COVID-19 literature [39].

## 6. PARTICIPATION

Teams submitted *runs* (synonymous with a 'submission') where a run consists of a sorted list of documents for each topic in the corpus and the document list for a topic is sorted by decreasing likelihood that the document is relevant to the topic (in the system's estimation). A TREC-COVID run was required to contain at least one and no more than 1000 documents per topic. TREC-COVID recognized three different types of runs: automatic, feedback, and manual. An automatic run is a run produced using no human intervention of any kind—the system is fed the test topics and creates the ranked lists that are then submitted as is. A manual run is a run produced with some human intervention, which may range from small tweaks of the query statement to multiple rounds of human search. A feedback run is automatic except in that it makes use of the (manually-produced) official TREC-COVD relevance judgments from previous rounds.

The list of participating teams and their corresponding number of submissions per round are shown in Table 4. Teams are listed by the team label provided by the team. No attempt was made to enforce consistency in this name, so the same team may be listed under separate rows for separate rounds. This means that, officially, 92 unique teams participated in TREC-COVID, but the real number may be somewhat less. In Rounds 1-4, up to 3 runs were allowed, whereas in Round 5 up to 8 runs were allowed. Most teams in Rounds 1-4 used the maximum allowable 3 runs (means: 2.53, 2.67, 2.55, 2.67), while in Round 5 only 6 of 28 teams submitted the maximum allowable 8 runs (mean: 4.5).

Table 4: Teams participating in all five TREC-COVID rounds, with run counts for each round. Rounds 1-4 limited participants to 3 runs. Round 5 limited participants to 8 runs.

| Team | Round 1 | Round 2 | Round 3 | Round 4 | Round 5 |
|---|---|---|---|---|---|
| 0_214_wyb | | | 2 | | |
| abccaba | 2 | | | | |
| anserini | | 2 | 3 | 3 | 8 |
| ASU_biomedical | | 3 | | | |
| AUEB_NLP_GROUP | | 1 | | | |
| azimiv | 1 | | | | |
| BBGhelani | 2 | 3 | | | |
| BioinformaticsUA | 3 | 3 | 3 | 3 | 6 |
| BITEM | 3 | 2 | 2 | 2 | |
| BRPHJ | | | | | 3 |

| Team | | | | | |
|---|---|---|---|---|---|
| BRPHJ_NLP | 3 | | | | |
| CincyMedIR | 3 | 3 | 3 | 3 | 8 |
| CIR | | | 3 | 3 | 2 |
| CMT | | 3 | | | |
| CogIR | | 3 | | | |
| columbia_university_dbmi | 2 | 2 | | | |
| cord19.vespa.ai | 1 | 2 | 3 | | |
| covidex | 3 | 3 | 3 | 3 | 8 |
| CovidSearch | | 3 | | | |
| CSIROmed | 3 | 3 | 3 | 3 | 3 |
| cuni | | 3 | | | |
| DA_IICT | 3 | | | | |
| DY_XD | | 3 | | | |
| Elhuyar_NLP_team | 3 | 3 | | | 5 |
| Emory_IRLab | | 2 | 2 | 3 | |
| Factum | 1 | 3 | 2 | | |
| fcavalier | | | | | 1 |
| GUIR_S2 | 3 | 3 | | | |
| HKPU | | 1 | | 3 | 8 |
| ielab | 3 | 3 | | | |
| ILPS_UvA | | | | 3 | |
| ims_unipd | | 3 | | | |
| IR_COVID19_CLE | 3 | 3 | | | |
| IRC | 3 | 2 | | | |
| IRIT_LSIS_FR | 2 | | 3 | | |
| IRIT_markers | 3 | 3 | | | |
| IRLabKU | 3 | 3 | 2 | | |
| ixa | 3 | | | | |
| julielab | 3 | | 3 | 3 | 1 |
| KAROTENE_SYNAPTIQ_UMBC | 3 | | | | |
| KoreaUniversity_DMIS | 3 | | | | |
| LTR_ESB_TEAM | | | 1 | | |
| MacEwan_Business | | | | | 1 |
| Marouane | | | | 2 | |
| MedDUTH_AthenaRC | | 3 | | | |
| mpiid5 | | 3 | 3 | 1 | 2 |
| NI_CCHMC | 3 | | | | |
| NTU_NMLab | 3 | | | | |
| OHSU | 3 | 3 | 3 | 3 | |
| PITT | | 3 | | | |
| PITTSCI | 3 | | | | |
| POZNAN | 3 | 3 | 3 | 3 | 3 |
| Random | | 1 | | | |
| req_rec | | 3 | | | |
| reSearch2vec | | | | | 7 |
| risklick | | 3 | 3 | 3 | 7 |
| RMITB | 2 | 1 | | | |
| RUIR | 3 | 3 | 1 | | |
| ruir | | | | | 3 |
| sabir | 3 | 3 | 3 | 3 | 8 |
| SavantX | 3 | 3 | | | |
| SFDC | 2 | 2 | 3 | 3 | 1 |
| shamra | | 1 | | | |
| Sinequa | 2 | | | | |
| Sinequa2 | 1 | | | | |
| smith | 3 | | | | |
| tcs_ilabs_gg | 1 | | | | |
| Technion | 3 | 3 | | | |
| test_uma | | | | 1 | |
| THUMSR | 3 | | | | |
| TM_IR_HITZ | 3 | | | | |
| TMACC_SeTA | 1 | 3 | | | |
| TU_Vienna | 2 | | | | |
| UAlbertaSearch | | | | 1 | 2 |
| UB_BW | 3 | 3 | 1 | | |
| UB_NLP | 1 | | | | |
| UCD_CS | | 3 | 3 | 3 | 3 |
| udel_fang | 3 | 3 | 3 | 3 | 3 |
| UH_UAQ | | | 1 | 2 | 7 |
| UIowaS | 3 | 3 | 3 | | 3 |
| UIUC_DMG | 3 | | | | |
| UMASS_CIIR | 2 | | | | |
| unipd.it | 3 | | | | |
| unique_ptr | 3 | 3 | 3 | 3 | 6 |
| uogTr | 3 | | 2 | 3 | 8 |
| UWMadison_iSchool | | | | 3 | |
| VATech | | | 3 | | |
| VirginiaTechHAT | 3 | 3 | | | |
| whitej_relevance | | 3 | | | |
| WiscIRLab | | | | | 6 |
| wistud | 3 | | | | |
| xj4wang | 1 | 3 | 3 | 3 | 3 |

# 7. POOLING

Relevance judgments are what turns a set of topics and documents into a retrieval test collection.  The judgments are the set of documents that should be returned for a topic and are used to compute evaluation scores of runs.  When the scores of two runs produced using the same test collection are compared, the system that produced the run with the higher score is assumed to be the better search system. Ideally we would have a judgment for every document in the corpus for every topic in the test set, but humans need to make these judgments (if the relevance of a document could be automatically determined then the information retrieval problem itself is solved), so a major design decision in constructing a collection is selecting which documents to show to a human annotator for each topic.  The goal of the selection process is to obtain a representative set of the relevant documents so that the score comparisons are fair for any pair of runs.

In general, the more judgments that can be obtained the more fair the collection will be, but judgment budgets are almost always determined by external resource limits. For TREC-COVID, the limiting factor was time.  Since the time between rounds was short, the amount of time available for relevance annotation was also short.  Based on previous TREC biomedical tracks, we estimated that we would be able to obtain approximately 100 judgments per topic per week with two weeks per TREC-COVID round, though that estimate proved to be somewhat low.

For most retrieval test collections, the number of relevant documents for a topic is very much smaller than the number of documents in the collection, small enough that the expected number of relevant documents found is zero when selecting documents to be judged uniformly at random while fitting within the judgment budget. But, search systems actively try to retrieve relevant documents at the top of their ranked lists, so the union of the set of top-ranked documents from many different runs should contain the majority of the relevant documents. This insight led to a process known as pooling that was first suggested by Spärck Jones and van Rijsbergen [40] and has been used to build the original TREC ad hoc collections. When scoring runs using relevance judgments produced through pooling, most IR evaluation measures treat a document that has no relevance judgment (because it was not shown to an annotator) as if it had been judged not relevant.

As implemented in TREC, pooling is performed by designating a subset of the submitted runs as *judged* runs and defining a cut-off level λ such that all documents retrieved at a rank ≤ λ in any judged run are included in the pool. With this implementation, different topics will have different pool sizes because pool size depends on the number of documents retrieved in common by the judged runs. Collection builders do not have fine control over the number of documents to be judged, but have gross control by changing the number of judged runs and/or changing the cut-off level λ. To fit within the judgment budget and time available for judging for each individual round of TREC-COVID, only some of the submitted runs were judged and λ was small. For example, for the first round of TREC-COVID, only one of the maximum of three runs per team was a judged run and λ=7.

While judged runs are only guaranteed to have λ documents per topic judged, and unjudged runs have no minimum guarantee at all, the premise of pooling is that in practice runs will have many more judged documents in their ranked lists since different runs generally retrieve many of the same documents, albeit in a different order. Figure 2 illustrates this effect for TREC-COVID submissions. The figure shows a box-and-whisker plot for the number of judged documents retrieved by a run to depth 50 for all runs submitted to a given round. The plotted statistics are computed over the number of topics in the round, and counts are based only on the judgment sets used to evaluate runs in that round. Different colors distinguish the judged and unjudged runs, light blue for judged runs and dark blue for unjudged runs. So, for example, in Round 1 where only 7 documents per topic were guaranteed to be judged, a sizeable majority of runs (both judged and unjudged) had a median value of more than twenty documents judged for a topic, and runs with the most overlap had medians of about 35/50 documents judged. The medians generally increased over the different TREC-COVID rounds. This was mainly caused by the submitted runs becoming more similar to one another as the rounds progressed, except for Round 5 where many more documents overall were judged since it was the last round which allowed for more judging time. There is a decrease in median number judged between Rounds 2 and 3. This dip is explained by the CORD-19 release used in Round 3 was much bigger than in Round 2 (see Table 1) so runs had both more room to diverge and significantly less training data for the new portion.

But what about runs that have little overlap with other runs and thus have relatively few judged documents to inform evaluation scores? Figure 2 shows that some runs with very little overlap with other runs were submitted to TREC-COVID. Even runs with relatively many judged documents can have unjudged documents at ranks important to the evaluation measure being used to score the run (for example, unjudged documents at ranks 8-10 when evaluating using Precision@10).  The default behavior of treating unjudged documents as if they were not relevant is a reasonable approximation if pools are sufficiently large to expect that most relevant documents have been found, but a simple counting argument demonstrates that shallow pools can find only a limited number of relevant documents. The question then becomes how shallow is too shallow, and there is no known way of answering that question without obtaining more judgments.  The individual rounds' judgment sets appear adequate for ranking the submissions made to the rounds[1], and the cumulative judgment set known as TREC-COVID Complete is much larger.  Researchers can easily detect the presence of unjudged documents in their own runs and decide how to proceed if detected. If the runs to be compared have similar numbers of unjudged documents, and especially if it is a small number of unjudged documents, then comparisons will be stable for a majority of measures. When the number of unjudged is skewed, it is best to take precautions such as using incompleteness-tolerant measures or requiring larger differences in scores before concluding that runs are actually different.

[INSERT FIGURE 2 HERE]

Figure 2: Number of documents judged in the top 50 ranks of a submission by round. The black line within a box is the median number of documents judged for that submission over the set of topics in that round.  Judged submissions (submissions that contributed to the qrels) are plotted in light blue and unjudged submissions are in dark blue.

## 8. ASSESSMENT

The goal of the assessment process is to manually label all of the pooled results for relevance to the corresponding topic. In TREC-COVID, each result could receive one of three possible judgment labels:
1. **Relevant**: the article is fully responsive to the information need as expressed by the topic, i.e. answers the Question in the topic. The article need not contain all information on the topic, but must, on its own, provide an answer to the question.

---

[1] https://ir.nist.gov/trec-covid/papers/rnd1runs_j0.5-2.0.pdf

2. **Partially Relevant**: the article answers part of the question but would need to be combined with other information to get a complete answer.
3. **Not Relevant**: everything else.

Performing the assessment requires a level of familiarity with biology and medicine, certainly above the level of the general population. As a result, individuals with specific skillsets needed to be recruited. The assessors generally came from three different groups. The first group was recruited from the MeSH indexers at the U.S. National Library of Medicine. Determining the relevancy of MeSH terms (essentially topics) to biomedical articles is the job of a MeSH indexer, so TREC-COVID assessment is a natural extension of their position. 17 indexers graciously agreed to assess up to 100 articles per week. The second group consisted of 10 OHSU medical students taking an elective, largely the result of the pandemic disrupting medical education and preventing them from taking part in clinical rotations. The third group was recruited from current and former students and postdocs at UTHealth, OHSU, and NLM, as well as the social networks of this group. All were required to have a medical degree or an appropriate biomedical science degree. With funding from AI2, we recruited 40 of these individuals to judge up to a maximum of 1000 articles each. While the indexers performed assessments throughout the entire project, the OHSU medical students primarily judged in the first few rounds, while the third group of assessors judged in the later rounds.

Before assigning topics, all assessors were asked for their preferences for judging individual topics, with the hope of aligning topics with expertise. Finally, while it is ideal in an IR evaluation to limit each topic to one assessor, the constraints of both timing and funding made this infeasible. However, to every extent possible assessors were assigned the same topic as prior rounds in order to minimize intra-topic disagreements. Double-assessment was not performed, as single-assessment has become standard in IR evaluations.

The web-based assessment platform used for TREC-COVID is shown in Figure 3. A URL corresponds to one topic for one assessor. For assessors assigned more than one topic, or for topics whose judgments needed to be split between multiple assessors in a round, multiple URLs were used. The assessor was provided with a list of articles to judge on the left, the topic information at the top of the page, and an iframe with the HTML/PDF of the article to be judged taking up most of the screen. No specific requirements were placed on the assessor (e.g., they did not have to read the entire article). It is assumed an assessor can judge 50 articles for a topic in one hour.

[INSERT FIGURE 3 HERE]
Figure 3. Assessment platform.

Figure 4 shows the number of judgments made for each topic, by round. As can be seen, an attempt was made to increase the number of judgments for later topics, so these were often pooled to a greater depth than the earlier topics. A consequence of pooling to the depth of each judged run, as opposed to some kind of depth across runs for a topic, is a fair degree of variability amongst the number of judgments per topic. In general, the greater the agreement between runs for a topic, the fewer articles were required to be judged. On the other hand, topics with sizable disagreement between runs meant a wider net needed to be cast to identify the relevant articles. Pooling to a specific depth on each run accomplishes both these goals. Figure 5 shows a different view of the per-round assessments by topic with the distributions of the assignments. It can clearly be seen how the first 30 topics intentionally had smaller pools so that the assessors could focus on the more recent topics, allowing their total number of judgments to catch up.

[INSERT FIGURE 4 HERE]
Figure 4: The number of articles judged per topic, by round.

[INSERT FIGURE 5 HERE]
Figure 5: Distributions of assignments per topic across rounds of judging.

Table 5 shows the number of judged documents in the final, cumulative qrels. It excludes articles that are not in the final version of CORD-19 and only the most recent judgment for an article if it was re-judged. The variation in the percent of relevant results is very high: topic 19 had only 7.9% of its judged articles considered relevant, while for topic 39 this was 77.3%. This is largely a reflection of the amount of information in CORD-19, though the difficulty of interpreting the topic and the differing standards of assessors certainly play a role as well. For reference, topic 19 is "*What type of hand sanitizer is needed to destroy Covid-19?*", while topic 39 is "*What is the mechanism of cytokine storm syndrome on the COVID-19?*".

Table 5: Counts of total numbers of judged documents and number of relevant documents per topic. Percent relevant is the fraction of judged documents that are some form of relevant.

| Topic | Total Judged | Partially Rel | Fully Rel | % Rel | Topic | Total Judged | Partially Rel | Fully Rel | % Rel |
|---|---|---|---|---|---|---|---|---|---|
| 1 | 1647 | 362 | 337 | 42.4 | 26 | 1720 | 148 | 684 | 48.4 |
| 2 | 1287 | 71 | 264 | 26.0 | 27 | 1477 | 580 | 321 | 61.0 |
| 3 | 1688 | 443 | 209 | 38.6 | 28 | 1103 | 74 | 543 | 55.9 |
| 4 | 1849 | 331 | 236 | 30.7 | 29 | 1241 | 275 | 374 | 52.3 |
| 5 | 1697 | 339 | 307 | 38.1 | 30 | 1035 | 211 | 193 | 39.0 |
| 6 | 1607 | 328 | 666 | 61.9 | 31 | 1701 | 213 | 158 | 21.8 |
| 7 | 1382 | 50 | 474 | 37.9 | 32 | 1571 | 80 | 149 | 14.6 |
| 8 | 1869 | 391 | 257 | 34.7 | 33 | 1270 | 125 | 182 | 24.2 |
| 9 | 1664 | 104 | 105 | 12.6 | 34 | 1842 | 74 | 124 | 10.7 |
| 10 | 1141 | 203 | 294 | 43.6 | 35 | 1360 | 32 | 207 | 17.6 |
| 11 | 1821 | 226 | 216 | 24.3 | 36 | 1233 | 105 | 572 | 54.9 |
| 12 | 1626 | 295 | 353 | 39.9 | 37 | 1234 | 144 | 369 | 41.6 |
| 13 | 1893 | 656 | 264 | 48.6 | 38 | 1920 | 618 | 765 | 72.0 |
| 14 | 1296 | 172 | 101 | 21.1 | 39 | 1264 | 438 | 539 | 77.3 |
| 15 | 1981 | 266 | 180 | 22.5 | 40 | 1230 | 217 | 371 | 47.8 |
| 16 | 1640 | 236 | 174 | 25.0 | 41 | 1043 | 87 | 269 | 34.1 |
| 17 | 1353 | 372 | 345 | 53.0 | 42 | 769 | 23 | 255 | 36.2 |
| 18 | 1325 | 319 | 347 | 50.3 | 43 | 878 | 97 | 203 | 34.2 |
| 19 | 1489 | 68 | 49 | 7.9 | 44 | 1238 | 182 | 360 | 43.8 |
| 20 | 1234 | 288 | 469 | 61.3 | 45 | 1171 | 352 | 549 | 76.9 |
| 21 | 1600 | 80 | 577 | 41.1 | 46 | 680 | 109 | 91 | 29.4 |
| 22 | 1325 | 216 | 379 | 44.9 | 47 | 1064 | 113 | 353 | 43.8 |
| 23 | 1293 | 194 | 201 | 30.5 | 48 | 747 | 202 | 279 | 64.4 |
| 24 | 1248 | 150 | 300 | 36.1 | 49 | 1093 | 131 | 136 | 24.4 |
| 25 | 1590 | 167 | 408 | 36.2 | 50 | 889 | 98 | 51 | 16.8 |

## 9. JUDGMENTS

The prior section described the manual assessment process. This section describes how those manual judgments are organized into distinct judgments sets to facilitate the evaluation of participant runs. After assessment is performed, the judgments are organized in files known as *qrels*. These are posted on the TREC-COVID web site. The format of an entry in a qrels is <topic-number,iteration,document-id,judgment> where topic-number designates the topic the judgments apply to, document-id is a CORD-19 document identifier, and judgment is 0 for not relevant, 1 for partially relevant, and 2 for fully relevant. The iteration field records the round in which the judgment was made. Annotators continued to make judgments on the weeks when participants were creating their runs for the next round, and judgments made during these weeks are labeled as "half round" judgments. That is, a document labeled as being judged in Round $X$.5 was selected to be judged from a run submitted to Round $X$ but was used to score runs submitted to Round $X$+1. For round 0.5, the documents were selected from runs produced by the organizers that are not official submissions. The judgment set for half Round $X$.5 was created by pooling runs submitted to Round $X$ deeper (i.e., using a larger value of λ) and/or adding to the set of judged runs. Documents that had been previously judged were removed from those pools.

Runs submitted to Round $X$ were scored using only the judgments made in judgment Rounds $X$-1.5 and $X$, not the cumulative set of judgments to that point. This was necessitated by the fact the relevance judgments from prior rounds were available to the participants at the time they created their submissions and they could use those judgments to create their runs (these were the feedback runs). To avoid the methodological misstep of using the same data as both training and test, TREC-COVID used *residual collection evaluation* [41] in all rounds after the first. In residual collection evaluation, any document that has already been judged for a topic is conceptually removed from the collection before scoring. Thus, participants were told not to include any previously judged documents in the ranked lists they submitted (even if that run did not make use of the judgments), and all pre-judged documents that were nonetheless submitted were automatically deleted from runs. The runs were then scored using the qrels built for that round. The runs that are archived on the web site are the runs as scored, that is, with all previously judged documents removed.

The combination of residual collection evaluation and a dynamic corpus results in a complicated structure. While later releases of CORD-19 are generally larger than earlier releases, later releases are not strict supersets of those earlier releases in that articles can be dropped from a release—because the article is no longer available from the original source or because the article no longer qualifies as being part of the collection according to CORD-19 construction processes, for example. Sometimes a "dropped" article has actually just been given a new document id, as can happen when a preprint is published and thus appears in a different venue. Document content can also be updated. For example, CORD-19 went through many changes between the May 1 and May 19 (TREC-COVID Rounds 2 and 3) releases. One result of these changes was that approximately 7000 articles were dropped between the two releases and approximately 600 of those dropped articles had been judged for relevance. Approximately 2000 of the 7000 dropped were articles whose document id had changed.

The valid use of a test collection to score runs requires that the qrels accurately reflect the document set. Documents that are no longer in the collection must be removed from the qrels because otherwise runs would be penalized for not retrieving phantom documents that are marked as relevant. Similarly, the qrels must use the correct document id for the version of the corpus regardless of which round the judgment was made in. Documents whose content was updated must be re-judged to see if the changed content makes a difference to the annotation. The naming scheme selected for the qrels reflects this complexity. The name of a TREC-COVID qrels file is composed of three parts, the header ("qrels-covid"); the document round (e.g., "d3"); and a range of judgment rounds (e.g., "j0.5-2"). The document round refers to the CORD-19 release that was used in the given TREC-COVID round, and all of the document ids in that file are with respect to that release.  The *TREC-COVID Complete* qrels is the cumulative qrels over all five rounds, with all document ids mapped to the July 16 release of CORD-19, using the document content as of the latest round in which the document was judged, and not including judgments for documents no longer in the collection.  Under the naming scheme, this qrels is "qrels-covid_d5_j0.5-5".  Note that because of residual collection evaluation, no TREC-COVID submission was scored using this qrels.  Round 5 runs were scored using "qrels-covid_d5_j4.5-5".

## 10. RESULTS OVERVIEW

The top five NDCG automatic/feedback runs (only the best run for each team) for each of the five rounds are shown in Table 6. Tables S1-S3 in the Supplemental Data contain the top 5 NDCG team runs for each of the three run submission types. More detailed per-round tables are available on the TREC-COVID website. Due to the depth of the pools, different rounds utilized different metrics.  Notably, Rounds 1-3 used P@5 and NDCG@10, while Rounds 4 and 5 used P@20 and NDCG@20. The table also lists which runs were included in the pooling process. Teams could select one of their runs per round to be judged. Since inferred measures were not used, runs that did not contribute to the pools are at a disadvantage. Most of the runs in Table 6 were judged, though this is likely a combination of the advantage given to a judged run and the fact that teams generally select what they believe to be their best run for judging.

Table 6. Top automatic/feedback runs (best run per team), as determined by NDCG, for each of the five rounds of TREC-COVID. P@*N*: Precision at rank *N*; NDCG@*N*: Normalized Discounted Cumulative Gain at rank *N*; MAP: Mean Average Precision; bpref: Binary Preference; judged?: whether the run contributed to the pooling.

**Round 1**

| team | run | runtype | P@5 | NDCG@10 | MAP | bpref | judged? |
|---|---|---|---|---|---|---|---|
| sabir | sab20.1.meta.docs | automatic | 0.7800 | 0.6080 | 0.3128 | 0.4832 | yes |
| GUIR_S2 | run2 | automatic | 0.6867 | 0.6032 | 0.2601 | 0.4177 | no |
| IRIT_markers | IRIT_marked_base | automatic | 0.7200 | 0.5880 | 0.2309 | 0.4198 | yes |
| CSIROmed | CSIROmedNIR | automatic | 0.6600 | 0.5875 | 0.2169 | 0.4066 | no |
| unipd.it | base.unipd.it | automatic | 0.7267 | 0.5720 | 0.2081 | 0.3782 | no |

**Round 2**

| team | run | type | P@5 | NDCG@10 | MAP | bpref | judged? |
|---|---|---|---|---|---|---|---|
| CMT | SparseDenseSciBert | feedback | 0.7600 | 0.6772 | 0.3115 | 0.5096 | yes |
| mpiid5 | mpiid5_run1 | feedback | 0.7771 | 0.6677 | 0.2946 | 0.4609 | no |
| UIowaS | UIowaS_Run3 | feedback | 0.7657 | 0.6382 | 0.2845 | 0.4867 | no |
| unique_ptr | UPrrf16lgbertd50-r2 | feedback | 0.7086 | 0.6320 | 0.3000 | 0.4414 | yes |
| GUIR_S2 | GUIR_S2_run2 | feedback | 0.7771 | 0.6286 | 0.2531 | 0.4067 | yes |

**Round 3**

| team | run | runtype | P@5 | NDCG@10 | MAP | bpref | judged? |
|---|---|---|---|---|---|---|---|
| covidex | covidex.r3.t5_lr | feedback | 0.8600 | 0.7740 | 0.3333 | 0.5543 | yes |
| BioinformaticsUA | BioInfo-run1 | feedback | 0.8650 | 0.7715 | 0.3188 | 0.5560 | yes |
| UIowaS | UIowaS_Rd3Borda | feedback | 0.8900 | 0.7658 | 0.3207 | 0.5778 | no |
| udel_fang | udel_fang_lambdarank | feedback | 0.8900 | 0.7567 | 0.3238 | 0.5764 | yes |
| CIR | sparse-dense-SBrr-2 | feedback | 0.8000 | 0.7272 | 0.3134 | 0.5419 | yes |

**Round 4**

| team | tag | type | P@20 | NDCG@20 | MAP | bpref | judged? |
|---|---|---|---|---|---|---|---|
| unique_ptr | UPrrf38rrf3-r4 | feedback | 0.8211 | 0.7843 | 0.4681 | 0.6801 | yes |
| covidex | covidex.r4.duot5.lr | feedback | 0.7967 | 0.7745 | 0.3846 | 0.5825 | yes |

| team | tag | type | P@20 | NDCG@20 | MAP | bpref | judged? |
|---|---|---|---|---|---|---|---|
| udel_fang | udel_fang_lambdarank | feedback | 0.7844 | 0.7534 | 0.3907 | 0.6161 | yes |
| CIR | run2_Crf_A_SciB_MAP | feedback | 0.7700 | 0.7470 | 0.4079 | 0.6292 | yes |
| mpiid5 | mpiid5_run1 | feedback | 0.7589 | 0.7391 | 0.3993 | 0.6132 | yes |

**Round 5**

| team | tag | type | P@20 | NDCG@20 | MAP | bpref | judged? |
|---|---|---|---|---|---|---|---|
| unique_ptr | UPrrf93-wt-r5 | feedback | 0.8760 | 0.8496 | 0.4718 | 0.6372 | yes |
| covidex | covidex.r5.2s.lr | feedback | 0.8460 | 0.8311 | 0.3922 | 0.533 | yes |
| Elhuyar_NLP_team | elhuyar_prf_nof99p | feedback | 0.8340 | 0.8116 | 0.4029 | 0.6091 | yes |
| risklick | rk_ir_trf_logit_rr | feedback | 0.8260 | 0.7956 | 0.3789 | 0.5659 | yes |
| udel_fang | udel_fang_ltr_split | feedback | 0.8270 | 0.7929 | 0.3682 | 0.5451 | yes |

Figure 6 shows the distribution of median scores for each topic by round. This empirically shows which topics are "easy" and "difficult", relatively speaking, based on system performance. If the topics were consistently easy or difficult across rounds, the marks for the given rounds would be in roughly the same order relative to other marks in that round. This is not the case, which suggests a variance of difficulty at ranking articles at medium ranks (since the later rounds are residual runs) as well as potential variability of the new articles in CORD-19 for that round. In a sense, this means the difficulty of a topic in a pandemic is in part relative to the time point at which that topic is queried.

Other trends can be observed in Figure 6 as well. Feedback runs outperform automatic runs, which makes sense as the feedback runs have access to topic-specific information to train their models. The median system also generally improved on a topic over the rounds. This applies both to feedback runs (which makes sense) and automatic runs (which is more surprising), though this could also be an artifact of the weaker teams dropping out of the challenge. A more detailed analysis of runs in Rounds 2 and 5 found that fine-tuning datasets with relevance judgments, MS-MARCO, and CORD-19 document vectors was associated with improved performance in Round 2 but not in Round 5 [8]. This analysis also noted that term expansion was associated with improvement in system performance, and that use of the narrative field in TREC-COVID queries was associated with decreased system performance.

[INSERT FIGURE 6 HERE]

Figure 6: Median average precision (AP) scores over all runs submitted to a given round. The topics on the x-axis are sorted by decreasing median AP.

As stated in the Introduction, a main motivation for pandemic IR is the ability to assess how methods can adapt to the needs of the pandemic as more information (both in the document collection and in manual judgments) become available. While we cannot conduct a detailed system-level analysis as we do not have access to the underlying systems, we can estimate the importance of relevance feedback relative to non-feedback systems (i.e., the automatic runs). In Round 2 (the first round for which feedback runs were possible), there were still 2 automatic runs in the top 10 (ranked by NDCG) and 9 automatic runs in the top 25. In Round 3, no automatic runs were in the top 10 and only 4 automatic runs were in the top 25. In Round 4, there were no automatic runs in the top 10, but the number in the top 25 increased to 9 runs. However, by Round 5, no automatic runs were in the top 25, and the best automatic run was ranked 33 by NDCG. This was not due to a lack of effort at developing non-feedback systems: there were 49 automatic runs submitted in Round 5 (39% of the total), and these were submitted by some of the top-performing teams from the feedback runs (covidex, unique_ptr, uogTr, etc.). Meanwhile, in Round 1 the top-performing automatic run (from sabir) utilized no machine learning (via transfer learning) or biomedical knowledge whatsoever. It has been remarked elsewhere that early in a pandemic, feature-rich systems still fail to outperform decades-old IR approaches [35]. The comparison of automatic versus feedback runs above, however, completes the spectrum to demonstrate that machine learning-based, feature rich systems do indeed outperform non-feedback based systems as the information about the pandemic increases.

## 11. METHODS OVERVIEW

In this section, we highlight the methods used by a handful of participants that have published papers or preprints on TREC-COVID. IR shared tasks are not well-suited to identifying a "best" method based solely on the ranking metrics from the prior section, and TREC historically has avoided referring to itself as a competition as well as declaring winners of a particular track. There are too many factors that go into a search engine's retrieval performance to empirically prove a given technique is better or worse just based on the system description provided by the authors. Further, a recent work attempts a comparative analysis of system features of the

TREC-COVID participants [8]. Instead, in this section we briefly focus on interesting aspects of TREC-COVID participants to illustrate the state of the field. Note that of the time of writing, most participants have not published (via preprint or peer review) a description of their system. What follows is the list of papers that have been reported to the organizers.

**SLEDGE** [42]. This automatic system used SciBERT [43] to re-rank the output of a BM25 retrieval stage. At least for Round 1, SLEDGE was trained on MS MARCO [44].

**CO-Search** [45]. This automatic system combined a question answering and abstractive summarization model to re-rank the output of a retrieval stage that utilized approximate k-nearest neighbor search over TF-IDF, BM25, and Siamese BERT [46] embeddings.

**NIR/RF/RFRR** [47]. This included a neural index run (NIR) automatic system that appended a BioBERT [48] embedding to the traditional document representation, an automatic relevance feedback (RF) system, and a relevance feedback with BERT-based re-ranking (RFRR) system.

**Covidex** [49]. This feedback system used T5 [50] to re-rank the output of a BM25 retrieval stage. A paragraph-level index was used instead of a document-level index.

**PARADE** [51]. This feedback system breaks documents into passages for special handling prior to using BERT [52] to re-rank the output of a BM25 retrieval stage.

**RRF102** [53]. This feedback system uses rank fusion to combine an ensemble of 102 runs. The constituent runs come from lexical and semantic retrieval systems, pre-trained and fine-tuned BERT rankers, and relevance feedback runs.

**Caos-19** [54]. This feedback system relied in a BM25 retrieval stage and added additional topic-relevant terms. These terms were based on Kaggle challenge tasks and WHO research goals.

## 12. LESSONS LEARNED

Here we organize a handful of the lessons learned in TREC-COVID. Most of what is described here has been discussed in some detail above, but we hope it is useful to organize it more succinctly here for emphasis. The lessons here are organized according to whether they were anticipated as well as the extent to which they were addressable during the course of the shared task. We follow this up with a set of recommendations for a future pandemic-like IR challenge, should the unfortunate need arise.

**Anticipated and Addressed**: Some major concerns were anticipated and ended up being well-addressed despite the sizable unknowns that still existed at the time TREC-COVID was launched.

First, our most immediate concern related to the logistics of manual assessment within the timeframes required to meet the goals of TREC-COVID. As Table 1 indicates, often there was less than 2 weeks to create judgments for roughly ten thousand pooled results. Section 8 describes the heterogeneous collection of assessors and funding used to conduct the manual assessment. It is clear that while this ended up being successful for TREC-COVID with over sixty nine thousand manual judgments, this is not a reliable model for future evaluations. It is possible that some sort of crowdsourcing of individuals with biomedical expertise may be a more reliable model, and is worthy of further investigation.

Second, unlike most other TREC tracks, TREC-COVID could not use the standard methodology of evaluating submissions using all judged documents. Because the judgments made for a round were publicly released after that round to support the use of relevance feedback, we needed to use an evaluation methodology that accounted for the training effect. Residual collection scoring is a traditional approach to feedback evaluation that is easy to understand and easy to implement, and it worked well in TREC-COVID. The most significant drawback to using residual collection scoring is that it forced all submissions to be scored over only a single round's judgments. As it turned out, the judgments from a single round were sufficient for stable comparisons among submissions (see more on this point below).

**Anticipated yet Not Problematic**: Next, some of our anticipated concerns ultimately ended up working out well, though not due to any specific effort on the part of the organizers.

First, we understood the judgment pools would likely be fairly shallow (that is, we would not identify the vast majority of relevant articles for each topic). This indeed ended up being the case, though not always for the reasons anticipated (see discussion of topics and document set below). The scientific problem with shallow pools is that they often lead to unstable estimates of system performance—systems that return a large percentage of top-ranked unjudged results cannot be fairly evaluated against those that consistently return results that are judged. We thus have differing levels of confidence in the run scores for each system. After a thorough evaluation of the stability of the collection in other work [9], this remarkably turns out not to be the case. Despite the shallow nature of the pools, it does not appear likely that judging significantly more results would have resulted in many changes in the system rankings provided in Section 10. While the system scores themselves would certainly be different (probably higher), the relative ranking would hardly change.

Second, the timeline of the task raised the concern about whether participants would be able to develop new approaches or adapt their existing systems. This applied both to having a system ready to participate in Round 1 as well as the ability to adapt systems between rounds. The participation numbers, including 56 teams in Round 1, indicate many researchers and developers were able to quickly deploy at least an initial technique for the task. The continued participation of many teams across rounds (roughly 30 teams in Rounds 3-5, though not the exact same teams) further suggests they were able to ingest the new data, re-train any models, and perform additional experiments during the short turnaround times for the task. This includes the use of state-of-the-art machine learning models (e.g., BERT, T5) that are well-known to require sizable compute loads. Of course, the short turnaround times did not give participants a sizable opportunity to experiment, but this is a realistic situation in a pandemic. As a result, it is perhaps quite a positive sign that most of the participant techniques heavily leveraged transfer learning as this may aid in the rapid response to future crises.

**Anticipated but Not Addressed**: At least one major challenge that we anticipated still remains an open question. A limitation of our evaluation is that it cannot really assess whether there are meaningful differences between runs. As can be seen from Table 6, there is not a strict correlation between metrics, which would suggest that even if there are statistically significant differences between runs on a metric, we do not know whether that metric is a good proxy for a user in this context. This would argue for the need for user studies for this domain to better calibrate metrics to actual search workflows. However, the best users for TREC-COVID to study were largely focused on COVID-19-related scientific inquiry and medicine, so conducting a user study in the middle of a pandemic would have been difficult.

**Not Anticipated**: Some challenges were entirely unanticipated prior to the launch of TREC-COVID. These largely stemmed from the nature of the information content in the document collection, including both its volume and velocity.

First, the quantity and granularity of topics proved a difficult challenge to manage. As described in Section 4, the topics were chosen through a variety of surveillance methods, and the 50 final topics did indeed reflect most of the key information needs of the pandemic (at least in terms of April-June 2020). And yet, the pace of the pandemic certainly resulted in more information needs emerging than the 5 new topics each round. From an evaluation standpoint, however, the real issue was the general nature of many of the topics. This resulted in hundreds of relevant documents for some topics (e.g., topic 38, "*COVID inflammatory response*", had 765 relevant results), which likely means that there are many more relevant documents that are unjudged. The impact of this is lessened by the stability assessment discussed above, but we would still suggest a different topic creation strategy for a future task. While there is no 'ideal' number of relevant documents for a topic for the purpose of IR evaluation, it is generally thought that having more topics—so long as they are nearly fully-judged—provides a fairer evaluation than fewer topics. We would thus recommend having more, but finer-grained topics. Further, our topic creation strategy did not involve extensive consultation with a wide body of experts (nor could it feasibly have done so, as described above), so coming up with a diverse set of realistic, fine-grained topics across the biological, clinical, and public health sciences would be quite difficult. But we still feel that, looking back, a larger number of finer-grained topics may well have represented a more realistic use case for expert users as well as a better evaluation for IR systems.

Second, an unexpected difficulty was the churn in the document set. We anticipated the document set would grow over the course of the pandemic, but the actual changes were much more significant. The overall document set did substantially grow in size over the course of the pandemic, but different versions of CORD-19 are not proper supersets of one another. Documents get dropped between rounds because they get withdrawn or change status (move from pre-print to published, for example) or no longer meet CORD-19's inclusion criteria. Some documents get new ids (are renamed) because of a status change. Documents that remain in the collection across two versions may contain different content in the versions. Any of the documents that changed or were dropped or renamed might have been judged in a prior round, complicating both the implementation of residual collection scoring and the definition of the judgment set for a given round. The complicated relationship among the judgment sets caused by this churn was the motivation for defining the different sets of relevance judgment files described in Section 13 and posted on the TREC-COVID website.

## 13. TEST COLLECTION

Lastly, we describe the different ways that TREC-COVID can be used as a test collection for IR research. Our goal is both to suggest different mechanisms for evaluation and to identify canonical benchmark tasks. These benchmark tasks include:

1. **TREC-COVID Complete**. This benchmark utilizes only the final version of CORD-19 used for the challenge (July 16 snapshot) and the Round 5 cumulative qrels file (with 69,318 judgments). This is the closest benchmark to a standard IR ad hoc task with a fixed corpus and no temporal component. It is suitable for automatic and manual approaches.

2. **TREC-COVID Chronological**. This benchmark utilizes the five cumulative qrels files, one for each round. The purpose of this benchmark is to evaluate automatic and manual approaches to assess their retrieval performance at different stages of the pandemic. Without any kind of learning, some systems may be more or less effective at different stages.

3. **TREC-COVID Chronological-ML**. This benchmark is similar to Chronological, but allows automatic and manual systems to train machine learning models on earlier topics for a round to evaluate on new topics. In Round 1, no topics are available for training, whereas in Round 5 all available judgments for the first 45 topics are available for training. This simulates the case where, over the course of a pandemic, manual labels are able to be gathered for existing topics but the focus is on predicting relevance only for new, unseen topics.

4. **TREC-COVID Residual**. This benchmark is for feedback systems that are allowed to train on prior rounds, but are only evaluated on the new judgments. This differs from the Chronological-ML benchmark in that both training is only allowed on the prior rounds and testing occurs on just the residual updates for that round. This simulates the case where certain "standing topics" get feedback over time and can be improved for future searchers.

As can be seen, these benchmarks evaluate different aspects of pandemic retrieval, with different benchmarks perhaps being suitable to different research communities.
The qrels that correspond to each of the above benchmarks are available on the TREC-COVID website, organized to be clear as to what judgments are available for training and testing for each round.

## 14. CONCLUSION

This paper described the TREC-COVID challenge, an IR shared task conducted in response to the COVID-19 pandemic and inspired by the need to develop search systems in an urgent, rapidly-evolving health crisis. The major goals of the challenge were to evaluate search engine performance for the COVID-19 scientific literature and to build a test collection for pandemic search. In terms of the evaluation, 92 unique teams submitted 556 runs based on manual, automatic, and feedback approaches. In terms of the test collection, we have described four different benchmark datasets based on the TREC-COVID judgments which will be useful for evaluating different perspectives on pandemic search. Overall, the task was extremely popular (exceeding the popularity of any prior TREC evaluation) and, despite the large logistical hurdles, was able to produce a large test collection for inspiring future research in pandemic search.

**Funding**

The Allen Institute of AI and Microsoft have contributed to funding.

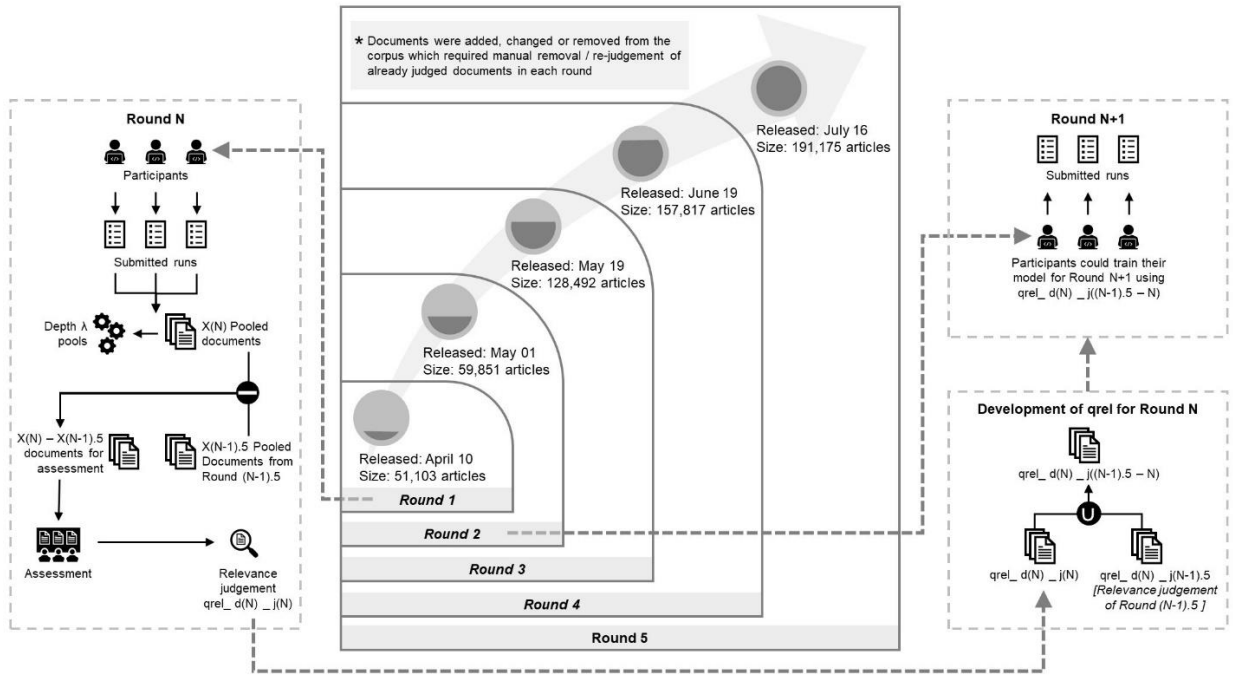

Figure 1. High-level structure of TREC-COVID.

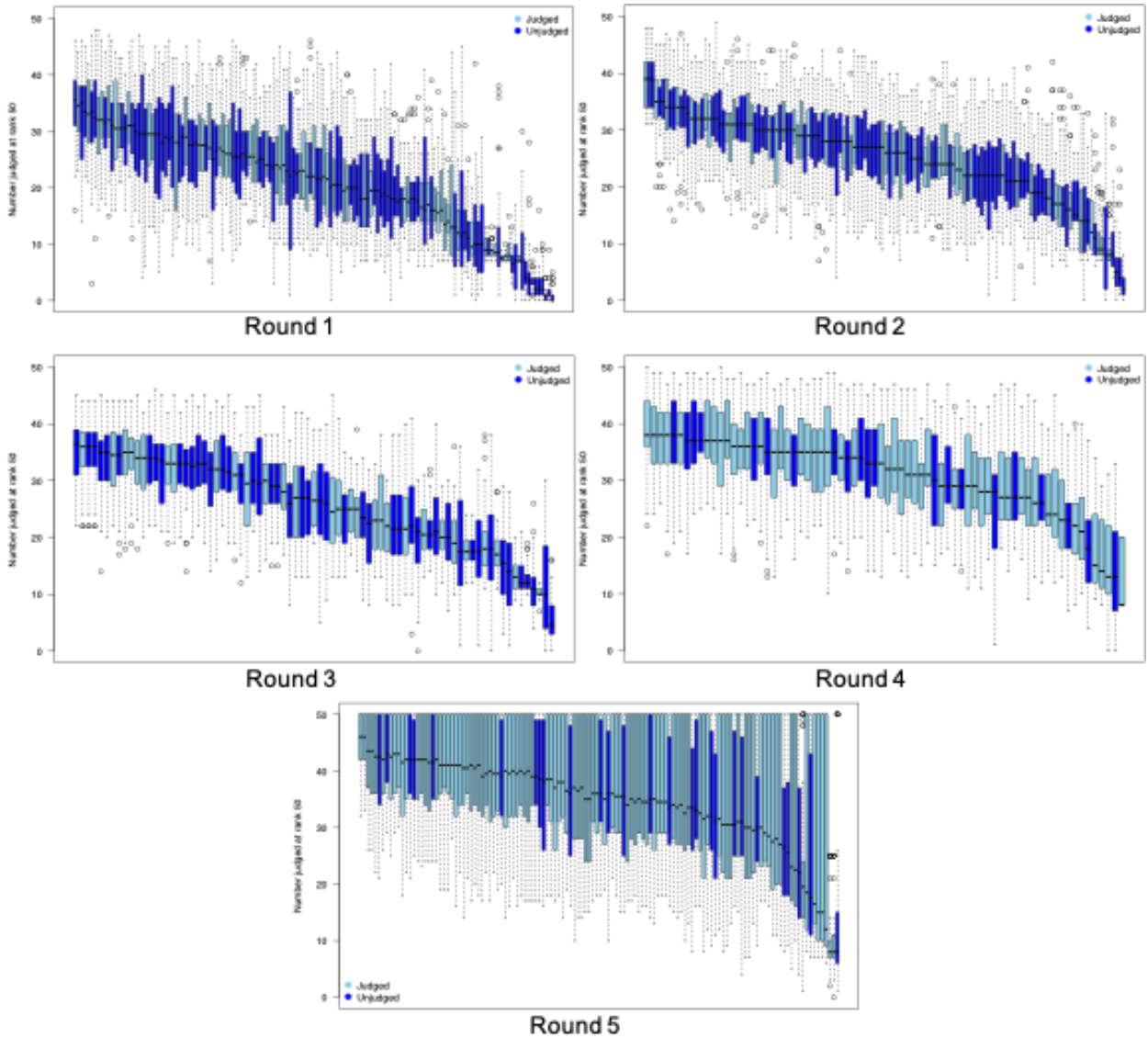

Figure 2: Number of documents judged in the top 50 ranks of a submission by round. The black line within a box is the median number of documents judged for that submission over the set of topics in that round. Judged submissions (submissions that contributed to the qrels) are plotted in light blue and unjudged submissions are in dark blue.

Figure 3. Assessment platform.

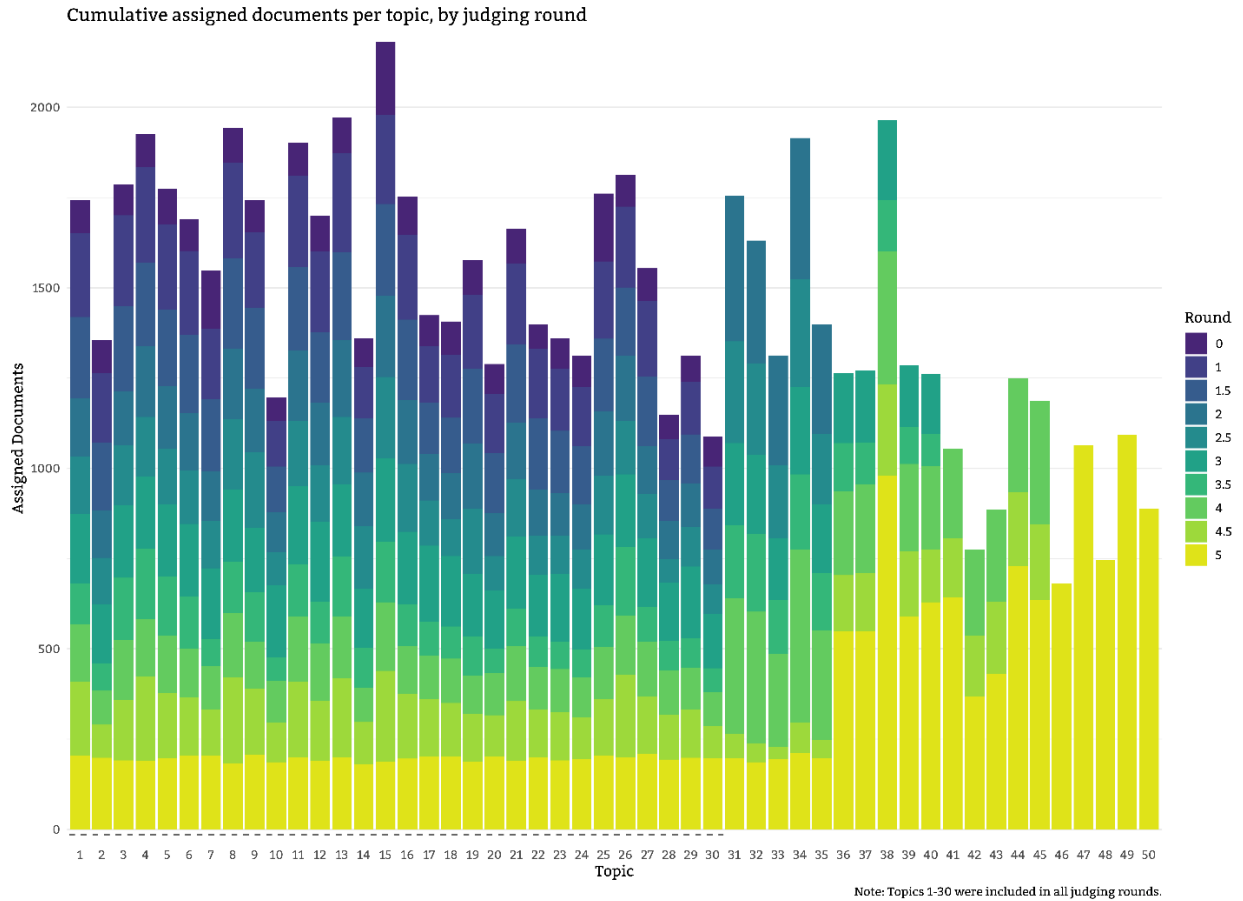

Figure 4: The number of articles judged per topic, by round.

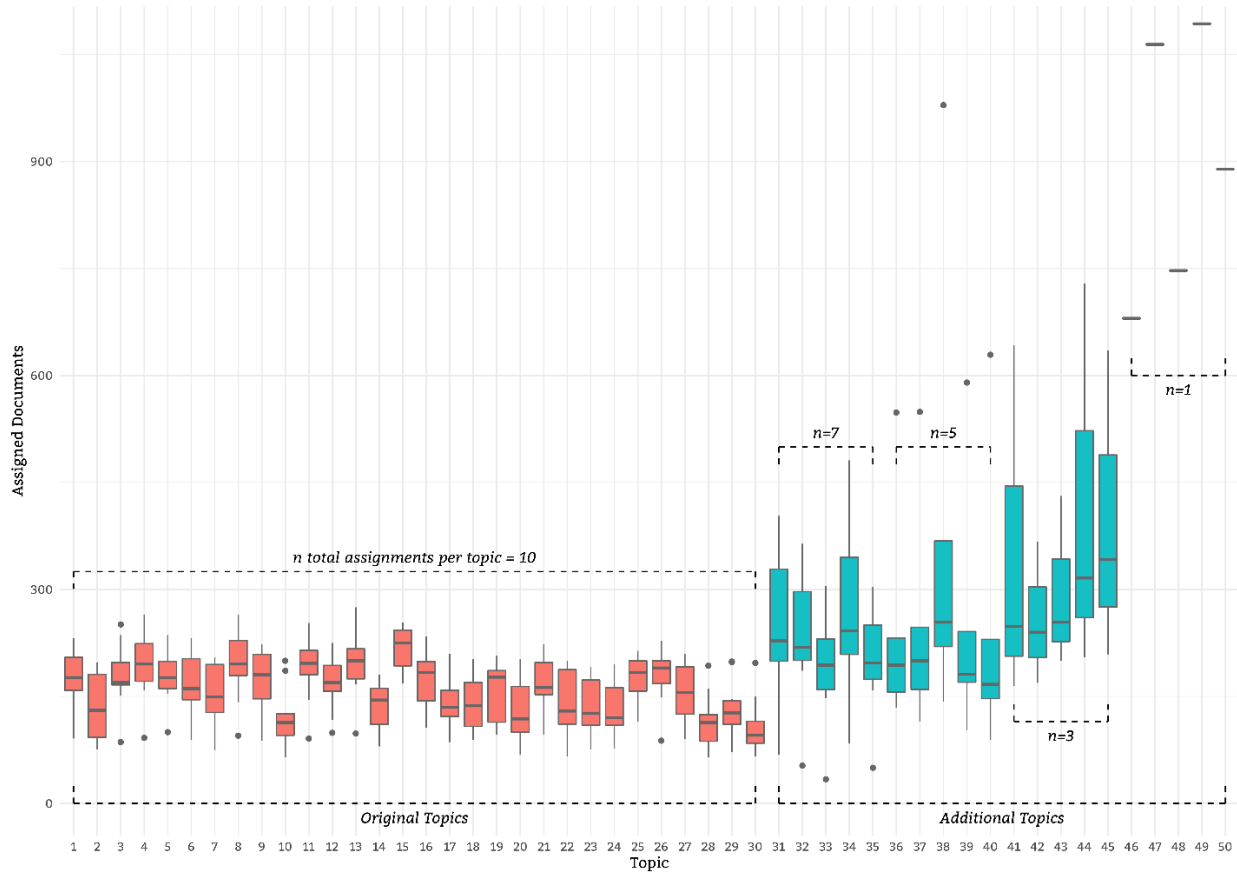

Figure 5: Distributions of assignments per topic across rounds of judging.

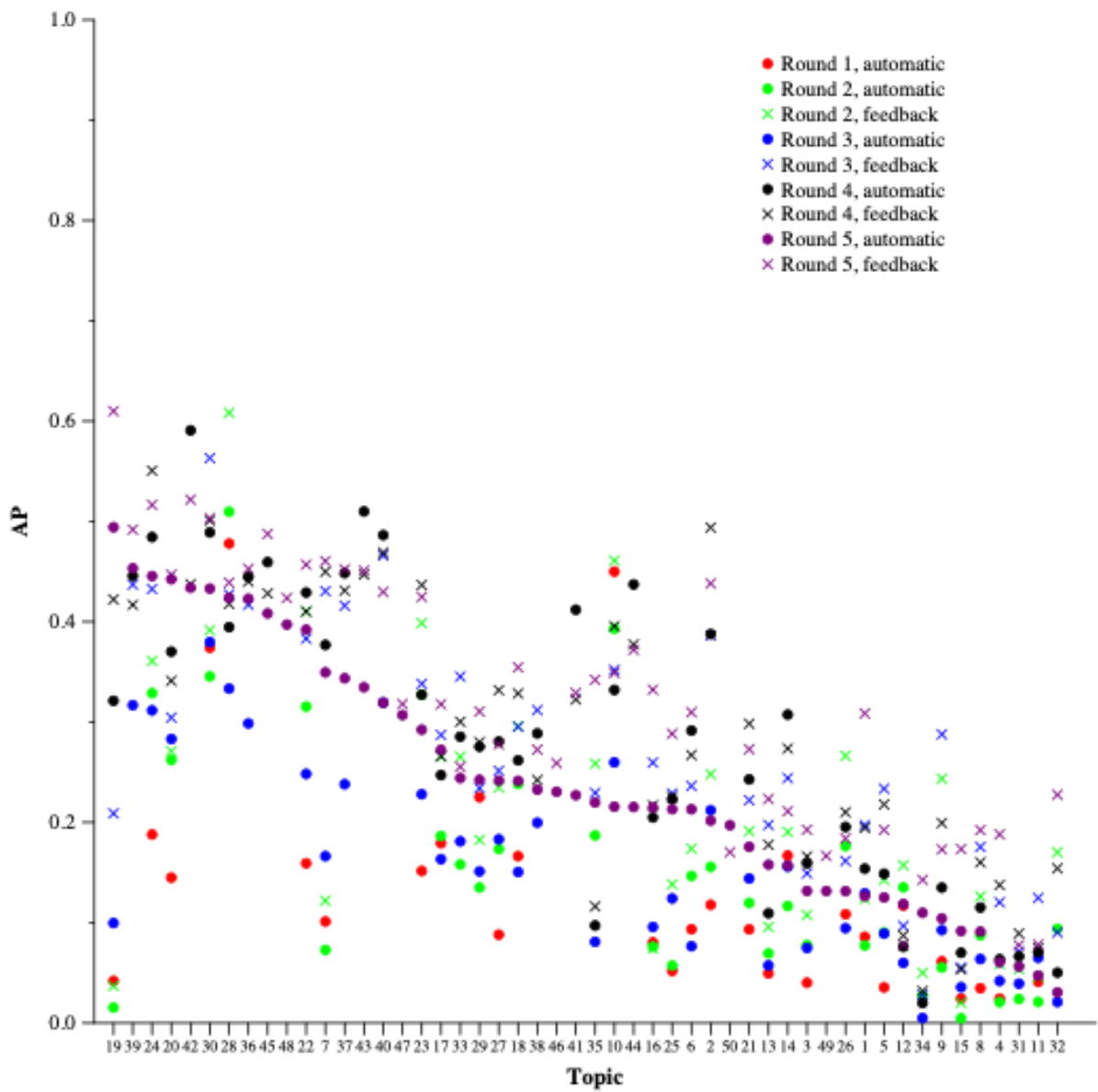

Figure 6: Median average precision (AP) scores over all runs submitted to a given round. The topics on the x-axis are sorted by decreasing median AP.

Table S1. Top 5 automatic runs (best run per team) as determined by the corresponding NDCG metric, for each of the five rounds of TREC-COVID. P@N: Precision at rank N; NDCG@N: Normalized Discounted Cumulative Gain at rank N; MAP: Mean Average Precision; bpref: Binary Preference; judged? whether the run contributed to the pooling. Full results available at https://ir.nist.gov/trec-covid/archive.html.

### Round 1

| team | run | P@5 | NDCG@10 | MAP | bpref | judged? |
|---|---|---|---|---|---|---|
| sabir | sab20.1.meta.docs | 0.7800 | 0.6080 | 0.3128 | 0.4832 | yes |
| GUIR_S2 | run2 | 0.6867 | 0.6032 | 0.2601 | 0.4177 | no |
| IRIT_markers | IRIT_marked_base | 0.7200 | 0.5880 | 0.2309 | 0.4198 | yes |
| CSIROmed | CSIROmedNIR | 0.6600 | 0.5875 | 0.2169 | 0.4066 | no |
| unipd.it | base.unipd.it | 0.7267 | 0.5720 | 0.2081 | 0.3782 | no |

### Round 2

| team | run | P@5 | NDCG@10 | MAP | bpref | judged? |
|---|---|---|---|---|---|---|
| GUIR_S2 | GUIR_S2_run1 | 0.7486 | 0.6251 | 0.2842 | 0.4569 | no |
| covidex | covidex.t5 | 0.7314 | 0.6250 | 0.2880 | 0.4876 | yes |
| CogIR | cogir-ibm-qQ-combs | 0.7086 | 0.6131 | 0.2590 | 0.4222 | no |
| Elhuyar_NLP_team | elhuyar_rRnk_cbert1 | 0.7200 | 0.5912 | 0.2941 | 0.4793 | yes |
| Emory_IRLab | Emory_IRLab_Run1 | 0.6971 | 0.5832 | 0.2247 | 0.3539 | yes |

### Round 3

| team | run | P@5 | NDCG@10 | MAP | bpref | judged? |
|---|---|---|---|---|---|---|
| SFDC | SFDC-fus12-enc23-tf3 | 0.7800 | 0.6867 | 0.3160 | 0.5660 | yes |
| covidex | covidex.r3.duot5 | 0.7700 | 0.6626 | 0.2676 | 0.5261 | no |
| IRLabKU | fusionofruns | 0.7350 | 0.6564 | 0.3084 | 0.5487 | yes |
| anserini | r3.fusion2 | 0.7150 | 0.6100 | 0.2641 | 0.4953 | no |
| sabir | sab20.3.metadocs_m | 0.6900 | 0.6032 | 0.2600 | 0.5202 | no |

### Round 4

| team | run | P@20 | NDCG@20 | MAP | bpref | judged? |
|---|---|---|---|---|---|---|
| covidex | covidex.r4.d2q.duot5 | 0.7267 | 0.7219 | 0.3122 | 0.5589 | yes |
| uogTr | uogTrDPH_QE_SCB1 | 0.7144 | 0.6820 | 0.3457 | 0.5510 | yes |
| unique_ptr | UPrrf38rrf2-r4 | 0.7156 | 0.6792 | 0.3612 | 0.5917 | no |
| SFDC | SFDC-re-fus12-enc24 | 0.7056 | 0.6763 | 0.3420 | 0.5874 | yes |
| UWMadison_iSchool | uw_base | 0.6867 | 0.6494 | 0.3471 | 0.5622 | no |

### Round 5

| team | run | P@20 | NDCG@20 | MAP | bpref | judged? |
|---|---|---|---|---|---|---|
| covidex | covidex.r5.d2q.2s | 0.7700 | 0.7539 | 0.3227 | 0.5052 | yes |
| uogTr | uogTrDPH_QE_SB_CB | 0.7910 | 0.7427 | 0.3305 | 0.4933 | yes |
| unique_ptr | UPrrf89-r5 | 0.7590 | 0.7235 | 0.3612 | 0.5364 | yes |
| udel_fang | udel_fang_nir | 0.7170 | 0.6830 | 0.3233 | 0.5555 | yes |
| sabir | sab20.5.metadocs_m | 0.7240 | 0.6778 | 0.3008 | 0.4874 | yes |

Table S2. Top 5 manual runs (best run per team) as determined by the corresponding NDCG metric, for each of the five rounds of TREC-COVID. P@N: Precision at rank N; NDCG@N: Normalized Discounted Cumulative Gain at rank N; MAP: Mean Average Precision; bpref: Binary Preference; judged? whether the run contributed to the pooling. Full results available at https://ir.nist.gov/trec-covid/archive.html.

### Round 1

| team | run | P@5 | NDCG@10 | MAP | bpref | judged? |
|---|---|---|---|---|---|---|
| xj4wang | xj4wang_run1 | 0.8333 | 0.6513 | 0.2367 | 0.4599 | yes |
| BBGhelani | BBGhelani1 | 0.8200 | 0.6689 | 0.3008 | 0.5294 | no |
| GUIR_S2 | run1 | 0.7933 | 0.6844 | 0.2663 | 0.4159 | yes |
| UMASS_CIIR | sheikh_bm25_manual | 0.7267 | 0.5790 | 0.1578 | 0.2095 | no |
| columbia_university_dbmi | cu_dbmi_bm25_1 | 0.7200 | 0.5887 | 0.2293 | 0.3605 | no |

### Round 2

| team | run | P@5 | NDCG@10 | MAP | bpref | judged? |
|---|---|---|---|---|---|---|
| mpiid5 | mpiid5_run3 | 0.8514 | 0.6893 | 0.3380 | 0.5679 | yes |
| OHSU | FullTxt_R2_Time | 0.7029 | 0.5969 | 0.2680 | 0.4525 | yes |
| xj4wang | xj4wang_run3 | 0.7314 | 0.5907 | 0.2210 | 0.4823 | no |
| BBGhelani | BBGhelani3 | 0.7543 | 0.5868 | 0.2386 | 0.5027 | no |
| columbia_university_dbmi | cu_dbmi_bm25 | 0.6171 | 0.5564 | 0.2083 | 0.4132 | no |

### Round 3

| team | run | P@5 | NDCG@10 | MAP | bpref | judged? |
|---|---|---|---|---|---|---|
| xj4wang | xj4wang_run1 | 0.8350 | 0.7431 | 0.2836 | 0.5681 | yes |
| risklick | active_learning | 0.8350 | 0.7285 | 0.2449 | 0.5065 | yes |
| OHSU | OHSU_Fusion | 0.7900 | 0.6816 | 0.3244 | 0.5828 | no |
| VATech | crowdPRF | 0.6950 | 0.6027 | 0.2320 | 0.4914 | no |

### Round 4

| team | run | P@20 | NDCG@20 | MAP | bpref | judged? |
|---|---|---|---|---|---|---|
| xj4wang | xj4wang_run1 | 0.7244 | 0.7019 | 0.2963 | 0.5507 | yes |
| UWMadison_iSchool | uw_crowd | 0.7189 | 0.6811 | 0.3923 | 0.6317 | yes |
| risklick | active_learning | 0.6533 | 0.6551 | 0.2946 | 0.5464 | yes |

### Round 5

| team | run | P@20 | NDCG@20 | MAP | bpref | judged? |
|---|---|---|---|---|---|---|
| WiscIRLab | uw_crowd1 | 0.7400 | 0.6877 | 0.3132 | 0.5036 | yes |
| xj4wang | xj4wang_run2 | 0.7030 | 0.6685 | 0.2509 | 0.5062 | yes |

Table S3. Top 5 feedback runs (best run per team) as determined by the corresponding NDCG metric, for Rounds 2-5 of TREC-COVID (feedback was not possible for Round 1). P@N: Precision at rank N; NDCG@N: Normalized Discounted Cumulative Gain at rank N; MAP: Mean Average Precision; bpref: Binary Preference; judged? whether the run contributed to the pooling. Full results available at https://ir.nist.gov/trec-covid/archive.html.

### Round 2

| team | run | P@5 | NDCG@10 | MAP | bpref | judged? |
|---|---|---|---|---|---|---|
| CMT | SparseDenseSciBert | 0.7600 | 0.6772 | 0.3115 | 0.5096 | yes |
| mpiid5 | mpiid5_run1 | 0.7771 | 0.6677 | 0.2946 | 0.4609 | no |
| UIowaS | UIowaS_Run3 | 0.7657 | 0.6382 | 0.2845 | 0.4867 | no |
| unique_ptr | UPrrf16lgbertd50-r2 | 0.7086 | 0.6320 | 0.3000 | 0.4414 | yes |
| GUIR_S2 | GUIR_S2_run2 | 0.7771 | 0.6286 | 0.2531 | 0.4067 | yes |

### Round 3

| team | run | P@5 | NDCG@10 | MAP | bpref | judged? |
|---|---|---|---|---|---|---|
| covidex | covidex.r3.t5_lr | 0.8600 | 0.7740 | 0.3333 | 0.5543 | yes |
| BioinformaticsUA | BioInfo-run1 | 0.8650 | 0.7715 | 0.3188 | 0.5560 | yes |
| UIowaS | UIowaS_Rd3Borda | 0.8900 | 0.7658 | 0.3207 | 0.5778 | no |
| udel_fang | udel_fang_lambdarank | 0.8900 | 0.7567 | 0.3238 | 0.5764 | yes |
| CIR | sparse-dense-SBrr-2 | 0.8000 | 0.7272 | 0.3134 | 0.5419 | yes |

### Round 4

| team | run | P@20 | NDCG@20 | MAP | bpref | judged? |
|---|---|---|---|---|---|---|
| unique_ptr | UPrrf38rrf3-r4 | 0.8211 | 0.7843 | 0.4681 | 0.6801 | yes |
| covidex | covidex.r4.duot5.lr | 0.7967 | 0.7745 | 0.3846 | 0.5825 | yes |
| udel_fang | udel_fang_lambdarank | 0.7844 | 0.7534 | 0.3907 | 0.6161 | yes |
| CIR | run2_Crf_A_SciB_MAP | 0.7700 | 0.7470 | 0.4079 | 0.6292 | yes |
| mpiid5 | mpiid5_run1 | 0.7589 | 0.7391 | 0.3993 | 0.6132 | yes |

### Round 5

| team | run | P@20 | NDCG@20 | MAP | bpref | judged? |
|---|---|---|---|---|---|---|
| unique_ptr | UPrrf93-wt-r5 | 0.8760 | 0.8496 | 0.4718 | 0.6372 | yes |
| covidex | covidex.r5.2s.lr | 0.8460 | 0.8311 | 0.3922 | 0.533 | yes |
| Elhuyar_NLP_team | elhuyar_prf_nof99p | 0.8340 | 0.8116 | 0.4029 | 0.6091 | yes |
| risklick | rk_ir_trf_logit_rr | 0.8260 | 0.7956 | 0.3789 | 0.5659 | yes |
| udel_fang | udel_fang_ltr_split | 0.8270 | 0.7929 | 0.3682 | 0.5451 | yes |